\documentclass[aps, prd, letterpaper, 12pt, nofootinbib, superscriptaddress, longbibliography, notitlepage]{revtex4-1}
\usepackage[utf8]{inputenc}
\usepackage{amsmath,amssymb,amsfonts}
\usepackage{mathrsfs}
\usepackage{color}
\usepackage{graphicx} 
\usepackage{ulem}
\usepackage[section]{placeins}
\usepackage[colorlinks=true,linkcolor=blue,citecolor=blue]{hyperref}
\allowdisplaybreaks

\pdfoutput=1 

\usepackage{graphicx}
\usepackage{hyperref}
\usepackage{aas_macros}
\usepackage{amsmath}
\usepackage{amssymb}
\usepackage{xcolor}
\usepackage{nccmath}
\usepackage{cleveref}

\def\beq{\begin{equation}\begin{aligned}}
\def\eeq{\end{aligned}\end{equation}}

\begin{document}

\title{\boldmath Dark Matter from quasi-de Sitter Horizons}

\author{Stefano Profumo,}
\email{profumo@ucsc.edu}
\affiliation{Department of Physics, 1156 High St., University of California Santa Cruz, Santa Cruz, CA 95064, USA}
\affiliation{Santa Cruz Institute for Particle Physics, 1156 High St., Santa Cruz, CA 95064, USA}

\begin{abstract}
\noindent Assuming that (1) the universe underwent a post-inflationary  accelerated expansion phase driven by a fluid with equation of state $P=w\rho$ and $-1<w<-1/3$,  that (2)  the cosmic horizon in an accelerating, quasi-de Sitter universe has a temperature inversely proportional to the proper size of the horizon, and that (3) we are static observers, we calculate the frozen-in density of a stable  particle of mass $m$ produced by the cosmic horizon that does not undergo any number-changing processes in the late universe. We find that, as a function of the equation of state and the temperature when radiation domination starts and the quasi-de Sitter phase ends, the mass of the dark matter producing the observed cosmological abundance via this mechanism ranges from 10 keV up to close to the Planck scale.
\end{abstract}

\maketitle
\newpage
\section{Introduction}
\label{sec:intro}

The phenomenon of particle production from de Sitter horizons, analogous to Hawking radiation from black holes, lies at a key intersection of quantum field theory, curved spacetime physics, and thermodynamics. This effect is fundamentally tied to the so-called Gibbons-Hawking mechanism, whereby a static observer in de Sitter spacetime experiences thermal radiation at the so-called Gibbons-Hawking temperature\footnote{Hereafter we use natural units.} $T_{\text{GH}} = H / (2\pi)$, determined by the Hubble constant $H$ (this phenomenon was {\it de facto} discovered in \cite{Figari:1975km}, and subsequently more clearly formulated and formalized in \cite{ref1}). 

Gibbons and Hawking's foundational work \cite{ref1} established and clarified the thermal nature of de Sitter horizon radiation and its thermodynamic properties, which include an entropy proportional to the horizon's area. Subsequent studies have confirmed and extended these early results and insights, emphasizing, in particular,  the importance of observer-dependent interpretations of vacuum states \cite{ref16}. For example, Bogoliubov transformations between in-vacuum and out-vacuum states consistently predict a thermal particle spectrum for static observers in de Sitter spacetime \cite{ref1,ref16,ref5,ref19}. These  results have been verified across a variety of settings, including applied quantum field theoretic treatments of massless and massive scalar fields \cite{ref8}, spinor fields \cite{ref8}, and semiclassical WKB tunneling calculations \cite{ref4,ref7}.

Modern approaches further extend this framework by incorporating dynamical effects and back-reaction. Studies have explored how particle production alters the background spacetime geometry, leading to a slow decay of the cosmological constant or modifications of the horizon \cite{ref3,ref10,ref23}. In particular, semiclassical tunneling mechanisms have been refined to account for these phenomena, showing that particle production in de Sitter spacetime slightly deviates from perfect thermality due to horizon dynamics \cite{ref13,ref4}. Additionally, alternative approaches, including Gaussian wave packets \cite{ref5}, phase-integral approximations \cite{ref6}, and super-adiabatic particle number formalism \cite{ref28}, provide corroborating insights into the way radiation emerges through quantum fluctuations near horizons.

The unerlying mechanism leading to the production of the cosmological dark matter (DM) is at present an open question and a matter of ongoing, intense scrutiny. While in many scenarios under consideration the DM arises as a thermal relic, either frozen out of frozen in by the background thermal bath \cite{Arcadi:2017kky, Arcadi:2024ukq}, other possibilities have been put forth, relevant for the present discussion. In particular, the possibility that the cosmological DM arises from the evaporation of light, primordial black holes (PBHs) in the early universe has been widely explored. The black hole Hawking temperature \( T_{\text{BH}} = \frac{1}{8\pi M} \), with $M$ the black hole mass, determines the radiation spectrum, with evaporation time scaling as \( t_{\text{ev}} \propto M^3 \). For PBHs with masses \( M \lesssim 10^{15} \) g, complete evaporation occurs before present times, making them potential DM sources \cite{Baldes:2021vyv, Kuehnel:2023rxt, Morrison:2018xla}. An asymmetric DM population can also arise in conjunction with the baryon asymmetry in the presence of CP-violating effective operators \cite{Smyth:2021lkn}. 
Recent studies also propose deviations from semiclassical evaporation. The ``memory burden'' effects, for instance, suppress radiation after PBHs lose half their mass, allowing lighter PBHs (\( M \sim 10^{10} \) g) to survive as DM candidates \cite{Kuehnel:2023rxt, Dvali:2024, Federico:2024fyt}. Modified gravity theories like scalar-tensor models, instead, predict faster evaporation rates, constraining PBH DM to heavier masses \cite{Dey:2024}.

In this study, in analogy to DM arising from black hole evaporation, we study the production of a massive, super-weakly (or non-) interacting DM species of a given mass from the quasi-de Sitter horizon of an early phase of the universe dominated by a species with equation of state $P=w\rho$ with $-1<w<1/3$; we assume that the universe transitions to radiation domination at some time well before the synthesis of light elements, i.e. above temperatures of around 10 MeV. Similar setups have been considered in the context of DM freeze-out \cite{Profumo:2003hq,DEramo:2017gpl} and freeze-in \cite{DEramo:2017ecx}. 
The goal of this study is to show in detail that, if we are static observers today, the observed cosmological DM, in a broad range of masses, may arise as a thermal product of a quasi-de Sitter cosmic horizon in the early universe, provided certain conditions are met for the reheating temperature,  the equation of state parameter $w$ driving the accelerated expansion phase, and the time at which the quasi-de Sitter phase ends. 

The remainder of this study is articulated as follows: The next  section describes and elucidates the notion of observer in de Sitter and quasi-de Sitter spacetimes. The following section \ref{sec:production} presents our assumptions and outlines our calculation of the DM abundance. The final section \ref{sec:conclusions} presents a discussion of our results and our conclusions.



\section{Observers in de Sitter and quasi-de Sitter spacetime}

Radiation from de Sitter horizons is inherently observer-dependent, due to the nature of quantum field theory in curved spacetime, where vacua are not universally defined but depend on the observer's position and motion relative to the horizon. In de Sitter spacetime, each observer experiences a distinct causal structure: {\it static observers} (those, that is, who remain at fixed spatial coordinates) are surrounded by a cosmological horizon due to the exponential space expansion, while {\it comoving observers} (following the Hubble flow) may perceive a different physical environment. As a result, phenomena like particle production depend on the observer's frame of reference and on their interaction with the quantum field modes. The dependence on the observer’s frame of reference is similar to the Unruh effect, where accelerated observers in Minkowski spacetime perceive a thermal spectrum while inertial observers do not. Similarly, static observers in de Sitter spacetime detect the horizon's radiation because they interact directly with the local vacuum structure through quantum mode evolution.

Note that the location and interpretation of the de Sitter horizon itself vary with coordinate choices. In static coordinates, the horizon behaves as a fixed causal boundary emitting thermal radiation, while in planar (comoving) coordinates, it appears as part of the expanding spacetime and is not thermally interpreted. This is again a reflection of the fact that the horizon’s thermodynamic effects, including radiation, depend on the observer’s ability to define the quantum field vacuum relative to the horizon.

Static observers  technically lie within the region covered by static coordinates, defined by $r < H^{-1}$, where $r$ is the radial coordinate in the metric:  
\begin{equation}
    ds^2 = -(1 - H^2 r^2) dt^2 + \frac{dr^2}{1 - H^2 r^2} + r^2 d\Omega^2,
\end{equation} and $H$ is the constant Hubble parameter of de Sitter spacetime \cite{GibbonsHawking1977, Parikh2002}. In this setting, an observer remains at a fixed spatial coordinate $r$ and measures the thermal radiation emanating from the cosmological event horizon. 

Can we assume that we are static observers? The notion of static observer and the possibility that we are such observers relies on several conditions. First, the observer must reside in a region of spacetime where static coordinates are well-defined, i.e., as noted above, within the static patch of de Sitter spacetime $(r < H^{-1})$ \cite{Parikh2002,Sekiwa2008}. This stands in contrast to global de Sitter coordinates or other metrics such as planar or FLRW (Friedmann-Lema\^itre-Robertson-Walker) coordinates, where the horizon does not appear stationary. Second, the spacetime must remain close to a maximally symmetric de Sitter solution, ensuring that the Hubble parameter, $H$, stays constant and the cosmological horizon remains static over time. This implies that dynamic effects—such as evolving $H$ during inflation or backreaction from particle production—are either absent or negligible. Under these assumptions, static observers perceive radiation arising from quantum field theory effects, which can be explained using Bogoliubov transformations that relate static observer mode functions to the global Bunch-Davies vacuum \cite{GibbonsHawking1977, Medved2002, Lapedes1978}. Tunneling methods, which treat particle emission as a quantum tunneling process from the horizon, also lend creedence to this perspective. Such methods  in fact derive a radiation spectrum that is nearly thermal under most circumstances \cite{Parikh2002, Medved2002, Sekiwa2008}.

{\color{black}The particle production mechanism proposed in this work, and detailed upon in the next section, shares conceptual ancestry with quantum field theoretic treatments of particle creation in time-dependent backgrounds, as developed in the seminal works of Parker~\cite{Parker:1968mvw, Parker:1969au}, Fulling~\cite{Fulling:1972md}, Hawking~\cite{Hawking:1974rv, Hawking:1975vcx}, and Gibbons and Hawking~\cite{Gibbons:1977mu}. In those approaches, field modes are globally quantized with respect to vacua defined at asymptotic past and future times. Particle production arises due to nontrivial Bogoliubov transformations between these in/out vacua, leading to a squeezed state interpretation of the evolving field. While this framework is well-established and rigorous, it typically requires knowledge of global spacetime structure and often focuses on idealized backgrounds such as exact de Sitter space or power-law inflation.

Our approach is complementary in spirit but operationally distinct. Rather than working with global vacuum modes, we consider local particle production associated with comoving modes as they cross the Hubble horizon. In this picture, particles are ``created'' when the physical wavelength of a comoving mode becomes comparable to the Hubble radius, and their phase-space distribution is determined by the scale-dependent redshifting governed by the evolving Hubble parameter $H(t)$. This yields a differential production rate that can be integrated over cosmological time, capturing contributions from multiple epochs.

A key difference lies in the physical interpretation of the produced particles. In standard QFT treatments, the particle concept is intrinsically observer-dependent and often tied to an adiabatic vacuum definition, leading to ambiguities in non-stationary backgrounds. In contrast, our framework assumes a well-defined particle species with mass $m$ and negligible interactions, allowing us to track their comoving number density and kinetic energy after horizon exit. This makes our approach particularly suited for modeling cold dark matter production, where the focus is not on particle number ambiguity, but on the emergence of a relic population with well-defined momentum and abundance. Our formalism naturally encodes redshifting effects and allows for continuous tracking of particle phase-space densities post-production. This facilitates a more direct computation of dark matter observables such as velocity dispersion and clustering properties, which are not easily extracted from standard QFT treatments. While both approaches aim to quantify particle creation in curved spacetimes, our horizon-based formulation emphasizes local, semiclassical, and cosmologically motivated features that are particularly relevant for scenarios involving cold, nonthermal relics such as dark matter.}

{\color{black} In discussing horizon-scale particle production, it is important to specify the class of observers with respect to which particle content is defined. While particle production in standard FRW cosmology is typically described from the point of view of comoving geodesic observers, effects analogous to the Gibbons–Hawking mechanism in de Sitter space suggest that a static observer—one associated with the apparent horizon—may provide a more natural frame for identifying a locally thermal particle flux. In our framework, the production of dark matter occurs well before the formation of local structures such as galaxies or the Earth, but the observer-dependence of particle content remains conceptually relevant. Our assumption is that the particles produced near the horizon can be meaningfully counted from the viewpoint of an effective observer anchored to the evolving causal structure, enabling a semiclassical description of their phase-space distribution that persists to late times.}



Let us now discuss constraints on an early quasi-de Sitter phase. Modified expansion histories prior to Big Bang Nucleosynthesis (BBN) include deviations from the standard cosmological framework driven by new energy components, scalar fields, or modified gravity. These modifications impact the Hubble expansion rate $H(T)$, affecting thermal relic abundances, freeze-out, and freeze-in processes. A faster-than-standard expansion rate, often modeled by additional energy components such as scalar fields whose energy density redshifts faster than radiation ($\rho_\phi \propto a^{-(4+n)}$ with $n > 0$), leads, for instance, to earlier freeze-out and higher relic abundances \cite{DEramo:2017ecx,DEramo:2017gpl,Visinelli:2017bny,Barman:2022qgt, Profumo:2003hq}. Phenomena like ``relentless DM," where annihilations persist past traditional freeze-out, may emerge in such cosmologies \cite{DEramo:2017ecx}. In contrast, slower expansion scenarios, such as those induced by early matter domination (EMD) from decaying particles, produce delayed freeze-out and reduced relic abundances while introducing entropy injection and non-thermal production effects \cite{Gelmini:2006pq,Maldonado:2019qmp,Allahverdi:2023kss}.

Scalar-tensor theories of gravity may also play a significant role in pre-BBN modifications. Enhanced or reduced Hubble rates are possible through couplings to scalar fields, leading to either higher relic densities (due to prolonged freeze-out) or extended supersymmetric parameter spaces \cite{Catena:2004ba,Catena:2007ix,Dutta:2016htz,Meehan:2015cna}. This underscores the sensitivity of thermal relics to the interplay between dark energy, gravitational modifications, and radiation transitions. Additionally, models with early scalar domination, such as those involving kination or quintessence, affect both freeze-out and freeze-in mechanisms, suppressing interactions and requiring larger annihilation cross-sections or couplings to match observed relic densities \cite{Profumo:2003hq,Dutta:2016htz,Barman:2022qgt,Okada:2004nc}.

Observational constraints, particularly from BBN and the Cosmic Microwave Background (CMB), impose strong constraints on these scenarios. Of special interest here, modified expansion models must transition to standard radiation dominance by $T \sim 10 \, \mathrm{MeV}$ to preserve light element abundances and ensure consistency with the effective number of relativistic species, $N_\text{eff}$ \cite{Gelmini:2006pq,Catena:2007ix,Meehan:2015cna}.

We also note that the transition from inflation to a second accelerated expansion phase in the very early universe can be achieved through several possible mechanisms: The inflaton field $\phi$ could evolve to a different region of its potential $V(\phi)$, where $-1 < w < -1/3$ \cite{Linde1982}; A separate field $\psi$ with a suitable potential $U(\psi)$ could become dominant, driving a late accelerated expansion \cite{Kofman1994}; another example is the model discussed in Ref.~\cite{Fichet:2023dju}, a holographic setup which can be thought either as a model of non-conformal strongly-interacting sector or as a non-AdS braneworld model. In either viewpoint, the model predicts the existence of an extra fluid in the 4D Friedmann equation with equation of state $-1 < w < -1/3$; finally, modifications to Einstein's field equations at high energies could induce accelerated expansion without requiring an additional field, see e.g. \cite{Starobinsky1980}. 
Each of these mechanisms could potentially lead to a phase of accelerated expansion with $-1 < w < -1/3$ immediately following inflation, at temperatures $T \gg 10$ MeV, before the onset of radiation domination.

 {\color{black} Here, both for simplicity and seeking to be model-independent, we just assume that the quasi-de Sitter phase with equation of state $-1 < w < -1/3$ ends in a standard reheating transition, during which the dominant vacuum-like energy density is converted into a thermal radiation bath. This process, while not modeled explicitly here, involves the decay or dissipation of the background component into relativistic particles and is accompanied by entropy production. We adopt the instantaneous reheating approximation, in which the energy transfer occurs rapidly at a characteristic temperature $T_r$, marking the onset of radiation domination. The entropy density $s(T_r)$ used to normalize the final dark matter yield reflects the entropy produced during this transition.
}


\section{Particle production from quasi-de Sitter horizons}\label{sec:production}

To compute particle production from a de Sitter or quasi de Sitter cosmic horizon, we integrate the black body\footnote{Utilizing the appropriate greybody factors is a small correction} emission of a (bosonic\footnote{This assumption is just for simplicity, the case of a fermionic particle produces almost identical numerical results.}) particle of mass $m$ from the thermal radiation off of the cosmic horizon. 
As indicated and explained above, we assume that at some point prior to BBN, conservatively, as also stated above, at temperatures $T>10$ MeV \cite{Gelmini:2006pq,Catena:2007ix,Meehan:2015cna}, the universe energy density is dominated by a species with equation of state $-1<w<-1/3$, thus leading to accelerated expansion and to the existence of a cosmic horizon. 

In quasi-de Sitter spacetimes, the cosmic horizon has a temperature inversely proportional to $L_H$, the proper size of the horizon, i.e. $T_{\rm GH}=1/(2\pi L_H)$; in turn, $L_H$ is related to the proper apparent horizon $ R_{AH}\equiv 1/H$ by
\begin{equation}
    \frac{R_{AH}}{L_H}= \frac{|1+3w|}{2},
\end{equation}
giving a temperature, for $-1<w<-1/3$, of
\begin{equation}
    T_w=\frac{H}{2\pi}\frac{|1+3w|}{2};
\end{equation}
As it should, $T_w\to H/(2\pi)$ in the pure de Sitter case $w\to 1$, and $T_w\to 0$ when the horizon disappears,  $w\to -1/3$.

\begin{figure}
\begin{center}
\includegraphics[width=0.65\columnwidth]{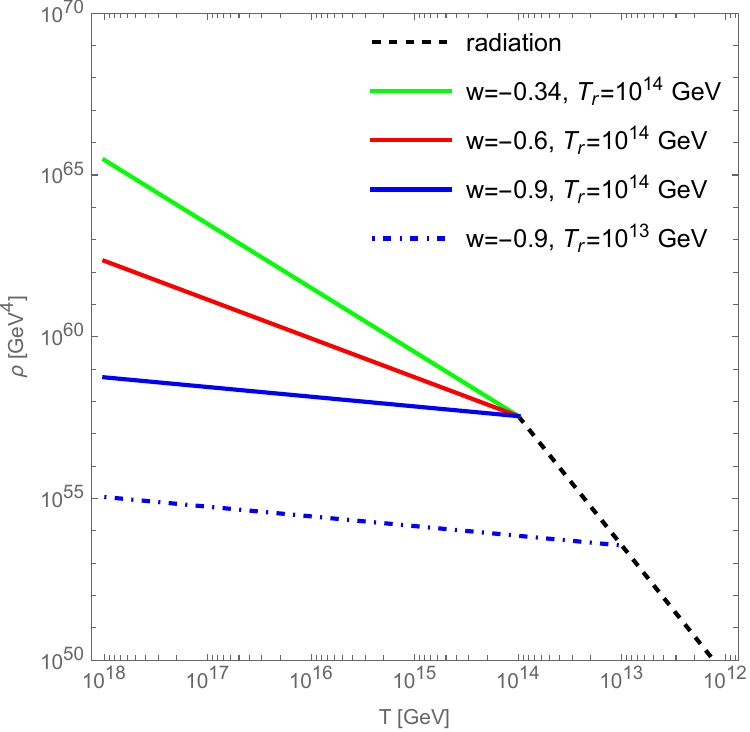}
\caption{Cumulative contribution to the universe's energy density of radiation (blue, right) and of fluids with three different equations of state parameters $w$ as a function of temperature. \textit{Note:} for $T > T_r$, the temperature axis should not be interpreted as a physical thermal bath temperature, but rather as a proxy for the inverse scale factor, $T \propto 1/a$, calibrated such that $T = T_r$ marks the onset of radiation domination.}
\label{fig:eos}
\end{center}
\end{figure}

For simplicity, we assume that the transition between radiation and the species or framework leading to the accelerated expansion driven by an equation of state with $-1<w<-1/3$ happens {\it instantaneously}, at a temperature $T_r>10$ MeV.  The Hubble rate at $T>T_r$ is then
\begin{equation}
H(T)=\frac{1}{\bar M_{\rm Pl}}\sqrt{\frac{\pi^2 g_*}{30} }\left(\frac{T}{T_r}\right)^{3(1+w)/2},
\end{equation}
with $g_*=106.75$ and $\bar M_{\rm Pl}\simeq 2.4\times 10^{18}$ GeV the reduced Planck mass. We show for clarity the assumed energy density structure, for a few choices of $w$ and $T_r$ in Fig.~\ref{fig:eos}. {\color{black} Note that for $T > T_r$, the temperature axis should not be interpreted as a physical thermal bath temperature, but rather as a proxy for the inverse scale factor, $T \propto 1/a$, calibrated such that $T = T_r$ marks the onset of radiation domination.}

In the geometric optics approximation, the differential emission rate for a bosonic particle from a horizon of temperature $T_w$ reads ($p$ here is the momentum of the particle):
\begin{equation}
    \frac{d^2N}{dp\ dt}=\frac{(27p^2)/(16H^2)}{e^\frac{4\pi\sqrt{p^2+m^2}}{|1+3w|H}-1}.
\end{equation}
The fermionic case has a ``+1'' in the denominator in the right hand side, and as noted above it yields essentially identical results as the exponential term in the denominator always dominates the integrals described below.

We intend to compute the comoving number density $Y=n/s$, where the entropy density $s= \frac{\pi^2}{45}g_{*}T^3$. This reads
\begin{equation}\label{eq:tintegral}
Y=\int_{T_i}^{T_f}dt(T)\int_0^\infty dp\ \frac{d^2N}{dpdt} \frac{1}{s(T)}\frac{1}{V_H},
\end{equation}
with $V_H=\frac{4\pi}{3}\frac{1}{H^3}$ the Hubble volume.
The quantity
\begin{equation}
dt(T)=\frac{dt}{dT}dT=\frac{d(1/H)}{dT}dT=-\frac{1}{H^2}\frac{dH}{dT}dT.
\end{equation}
Our final expression is thus
\begin{eqnarray}\label{eq:integrand}
\nonumber Y&=&\int dp\int dT \frac{27 p^2}{16 H^2}\frac{3H^3}{4\pi s(T)}\frac{1}{H^2}\frac{dH}{dT}\frac{1}{e^\frac{4\pi\sqrt{p^2+m^2}}{|1+3w|H}-1}\\
&=&\int dp\int dT \frac{81 p^2}{64\pi H(T) s(T)}\frac{dH(T)/dT}{e^\frac{4\pi\sqrt{p^2+m^2}}{|1+3w|H}-1}.
\end{eqnarray}

The extrema of integration for momentum depend on what is assumed about wavenumbers exceeding the horizon size, as well as on the largest temperature at which the species driving the de Sitter phase dominates the energy density of the universe, which is essentially the reheating temperature $T_{\rm reh}$ at the end of the inflationary period. 

\begin{figure}
\begin{center}
\mbox{\includegraphics[width=0.5\columnwidth]{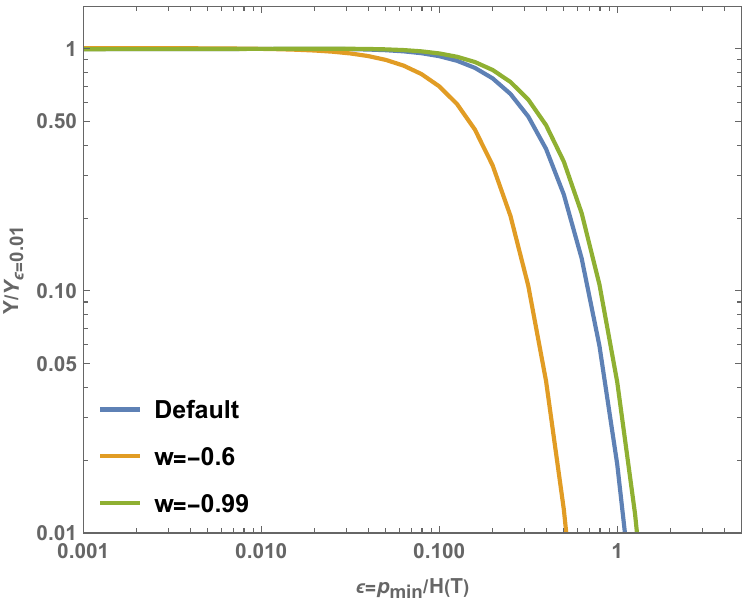}\quad \includegraphics[width=0.50\columnwidth]{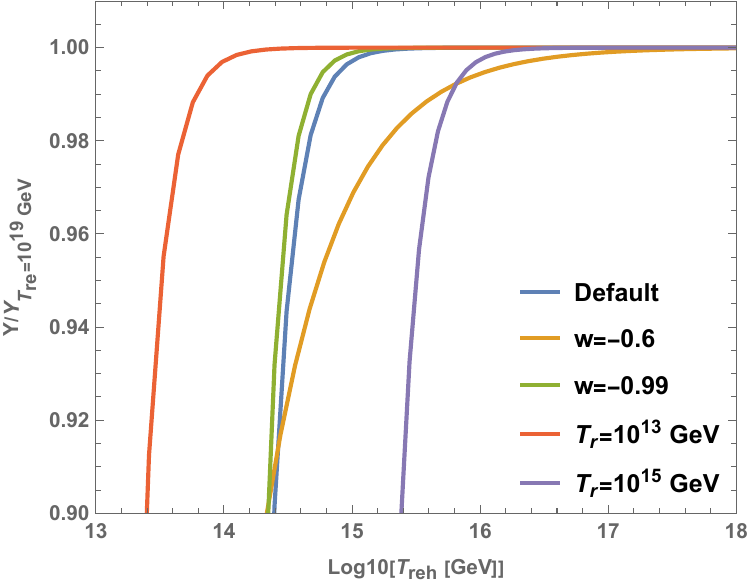}}
\caption{Effects of the integration limits choices on the abundance of frozen-in DM from the cosmic horizon. On the left we study the impact of the choice of the lower-limit of interaction in the momentum integral; our default choice is $\epsilon=0.01$; on the right we highlight the dependence of the upper limit of integration on the reheating temperature $T_{\rm reh}$}
\label{fig:limitintegrand}
\end{center}
\end{figure}

The choice of the lower integration limit $p_{\rm min}$ can be expressed, as a function of temperature, by the ratio $\epsilon=p_{\rm min}/H(T)$. Effectively, $\epsilon$ indicates the cutoff to momenta; for $p\gg H$, this corresponds to the physically motivated choice of requiring that the DM de Broglie wavelength $\lambda=1/p$ be much smaller than the cosmic horizon size $1/H$. 
Note that the integrand drops quickly for $p\gg H$, making the upper limit of integration in momentum immaterial. 

In Fig.~\ref{fig:limitintegrand} we study the effect of the variation of what numerically we find to be the only parameter that affects the abundance of the frozen-in DM, the equation of state parameter $w$, for   a ``Default''  choice for the remaining  parameters of $T_r=10^{14}$ GeV, $w=-0.9$, and $m=10^6$ GeV (the latter parameter has no influence on these plots). Hereafter we utilize $\epsilon=0.01$, and the figure shows the ratio of the DM yield for a given $\epsilon$ and for $\epsilon=0.01$; it is only for very large choices of $p_{\rm min}$, well into the expected exponential suppression, that we find any significant deviation from the $\epsilon=0.01$ case. The latter is notable especially for low $|w|$, i.e. further away from the de Sitter limit.

 {\color{black} It is important to note that during the quasi-de Sitter epoch preceding reheating, the universe is not in thermal equilibrium and the usual concept of entropy density $s = \frac{2\pi^2}{45} g_* T^3$ does not apply in a strict sense. The background energy density is dominated by a vacuum-like component with negligible radiation. Therefore, temperatures $T > T_r$ appearing in the integrand of Eq.~\eqref{eq:tintegral} should not be interpreted as physical radiation temperatures, but rather as redshift markers tied to comoving scales via $T \propto 1/a$, normalized to the post-reheating radiation bath.

We use the entropy density $s$ only at or below the reheating temperature $T_r$ to compute the final yield. In this context, $s(T_r)$ reflects the standard radiation-dominated value after reheating is complete. The appearance of $T$ in the integral is thus a bookkeeping device, not an assertion of thermal equilibrium prior to reheating\footnote{This approach is consistent with other analyses in which particle production occurs before reheating, such as gravitational production of dark matter in inflationary cosmologies (e.g.,~\cite{Chung:1998zb, Garny:2015sjg}).}.}

For the temperature integral, the upper limit of integration is naturally set by a reheating scale describing the effective end of the inflationary period and the (instantaneous) beginning of the second, quasi-de Sitter accelerated expansion phase when the DM is frozen in. We indicate this temperature with $T_{\rm reh}$. The right panel of Fig.~\ref{fig:limitintegrand} shows the ratio between the DM yield for a given $T_{\rm reh}$ to the ``asymptotic'' choice $T_{\rm reh}= M_P\sim 10^{19}$ GeV, with $M_P$ the Planck scale, as a function of  $T_{\rm reh}$. We use, as a ``Default'', the choices $T_r=10^{14}$ GeV, $w=-0.9$, and $m=10^6$ GeV (the latter parameter has no influence on this plot), and vary both $w$ and $T_r$. We find that the particular choice of $T_{\rm reh}$ is only relevant for large $T_r$ (as expected, since for $T_r\to T_{\rm reh}$ the yield vanishes), and for smaller $|w|$, i.e., again, the limit of largest-possible departure from  de Sitter.

\begin{figure}[!h]
\begin{center}
\hspace*{-0.5cm}\mbox{\includegraphics[width=0.5\columnwidth]{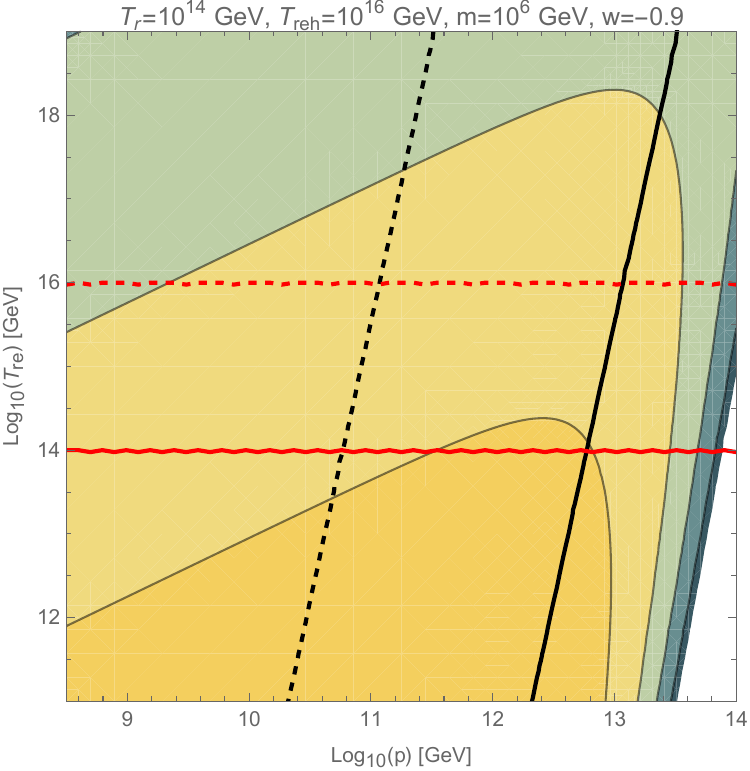}\quad\includegraphics[width=0.5\columnwidth]{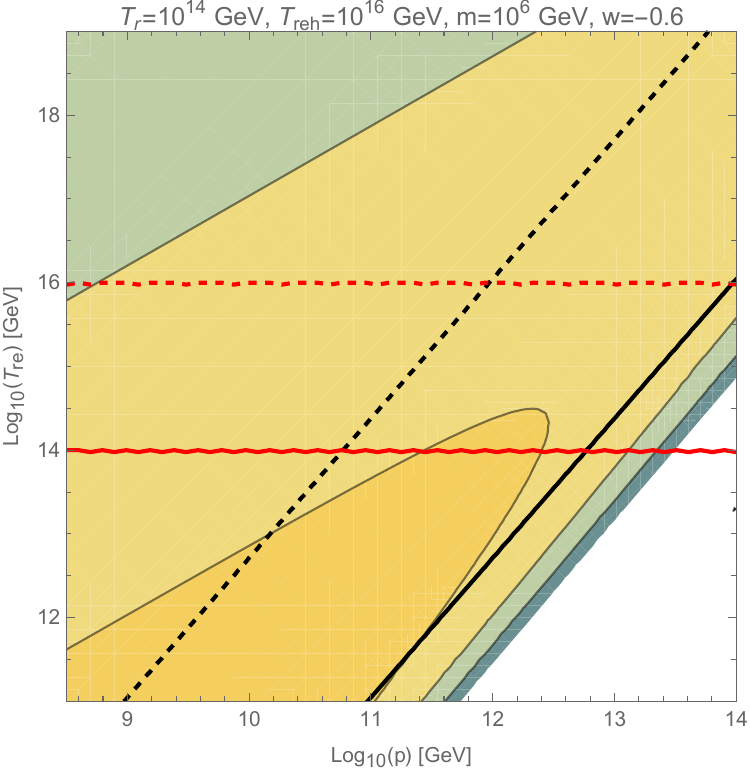}\includegraphics[width=1.2cm]{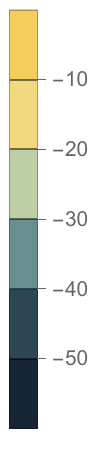}}\\[1.5cm]
\hspace*{-0.5cm}\mbox{\includegraphics[width=0.5\columnwidth]{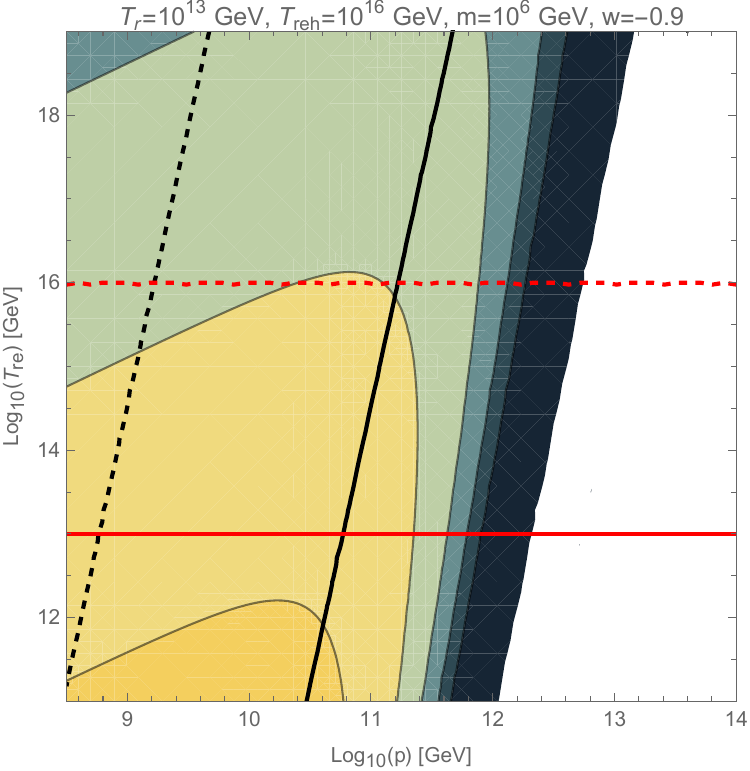}\quad\includegraphics[width=0.5\columnwidth]{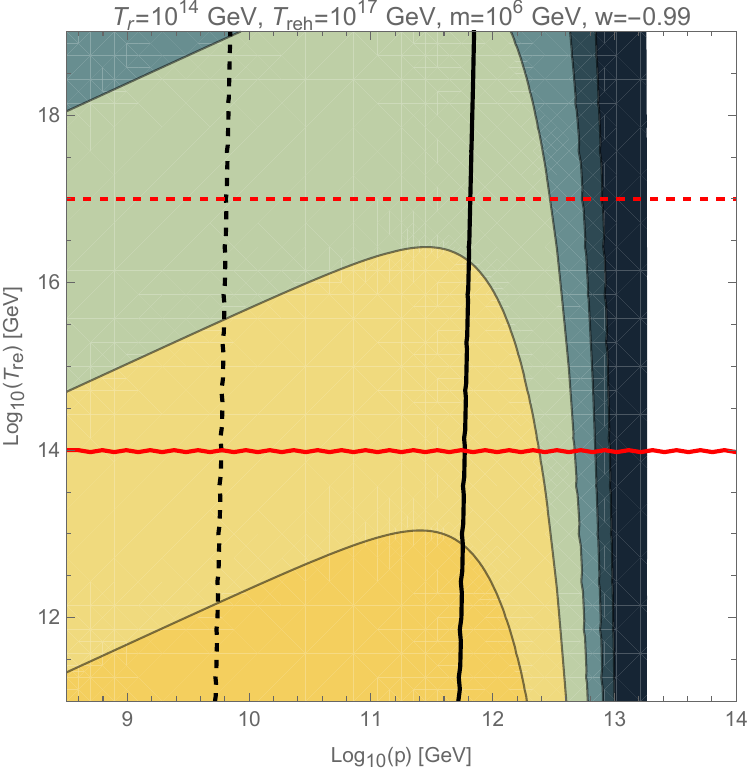}}
\caption{The integrand in Eq.~(\ref{eq:integrand}), in units of GeV$^{-2}$, for a number of choices of $T_r$, $T_{\rm reh}$, $m$, and $w$; the color scale indicates the Log10 of the value of the integrand, in units of GeV$^{-2}$.}
\label{fig:integrand}
\end{center}
\end{figure}

We show the integrand, multiplied by the integration variables $p$ and $T$, in Fig.~\ref{fig:integrand}, for a few choices of $T_{\rm reh}$ departing from our default choice of $T_{\rm reh}=10^{16}$ GeV, for different values of $w$, and of $T_r$.  
In the figures, the lower horizontal red line indicates $T_{r}$, the lower limit of integration in temperature (under our assumptions, there is no horizon at lower temperatures) and the upper, dashed red line the choice of $T_{\rm reh}$ (in other words, the integral in temperature is in between the two red lines). We also show the contours corresponding to $p=H(T)$ (solid black line) and $p=0.01 H(T)$ (dashed line): the integral in momentum runs to the left of the black lines, depending on the choice of $\epsilon=p/H(T)$, and is effectively cut off by the Boltzmann exponential suppression at large $p$.

\begin{figure}
\begin{center}
\mbox{\includegraphics[width=0.5\columnwidth]{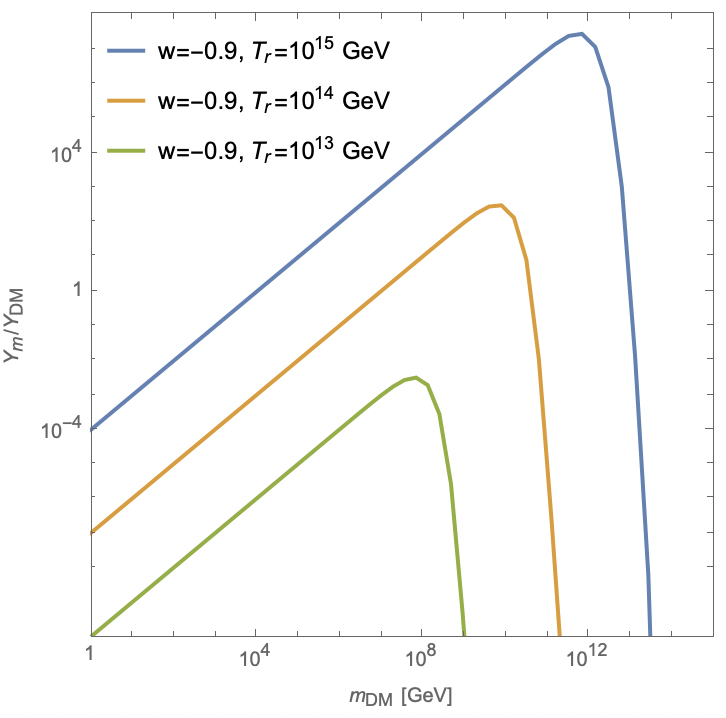}\qquad \includegraphics[width=0.50\columnwidth]{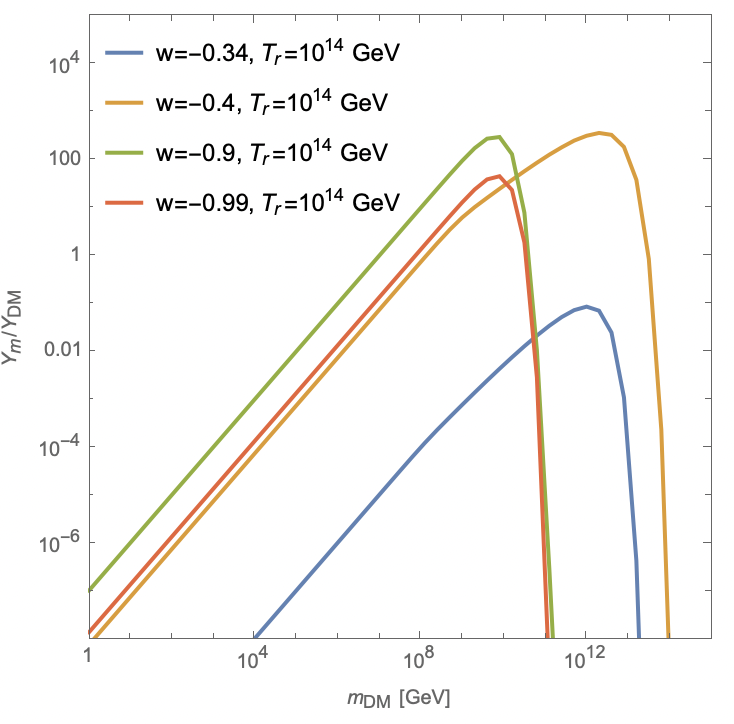}}
\caption{DM yield from the cosmic horizon as a function of mass, for a few values of $T_{\rm reh}$ and the equation of state $w$.}
\label{fig:yield}
\end{center}
\end{figure}

\begin{figure}
\begin{center}
\mbox{\includegraphics[width=0.45\columnwidth]{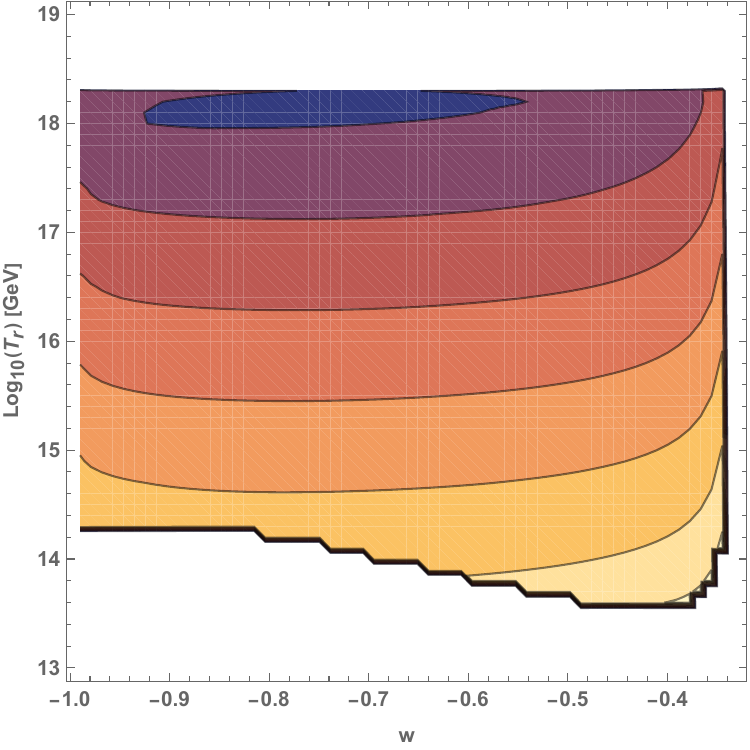}\includegraphics[width=0.14\columnwidth]{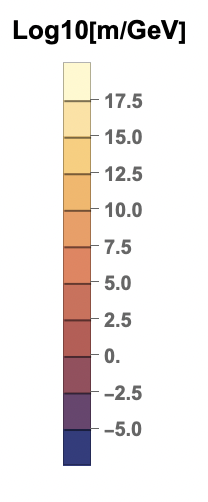}\includegraphics[width=0.45\columnwidth]{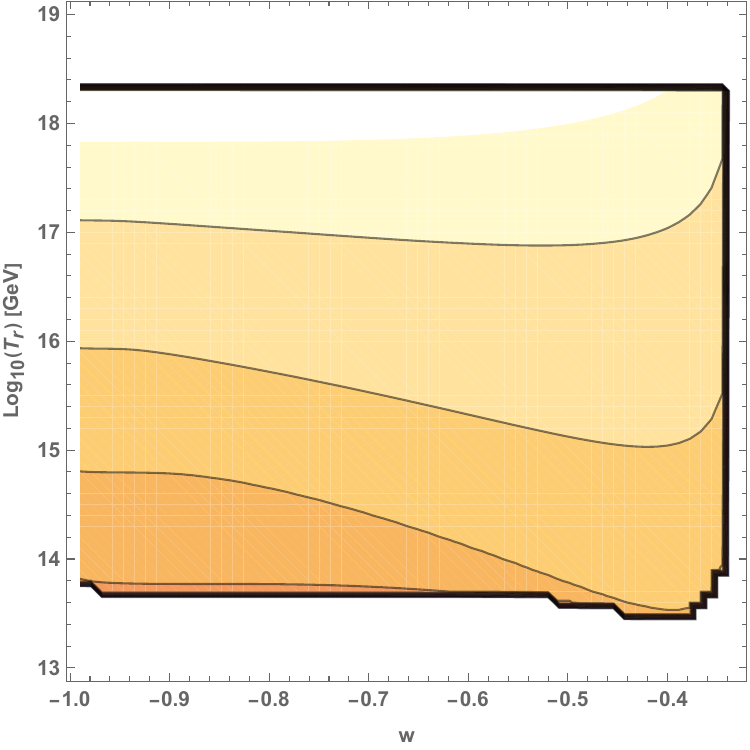}}
\caption{The lower and upper mass solutions producing the observed DM abundance on the $(w,T_r)$ plane. The contours show the Log10 of the corresponding DM mass in units of GeV. There are no solutions in the white upper and lower regions of the two plots}
\label{fig:yieldcontour}
\end{center}
\end{figure}

The scenario under consideration here can thus be viewed as possessing three independent parameters: the DM mass $m$, the equation of state parameter $w$, and the transition  temperature $T_r$, with a minimal dependence on the choices of the lower limit of integration in momentum $\epsilon$ and on the temperature corresponding to the start of the quasi de Sitter phase $T_{\rm reh}$ as discussed above. We argue that there exist viable particle masses $m$ for almost any $(w,T_r)$ choice such that the frozen-in DM yield matches observations. Figures ~\ref{fig:yield} and \ref{fig:yieldcontour} illustrate this point. 

In Fig.~\ref{fig:yield} we show, assuming $T_{\rm reh}=10^{18}$ GeV, thus close but lower than the Planck scale (again, the dependence of the DM yield on this parameter is illustrated in the right panel of Fig.~\ref{fig:limitintegrand}) the ratio between the yield, expressed as the comoving number density $Y=n/s$, and the observed comoving DM abundance $Y_{\rm DM}\simeq0.44\ {\rm eV}/m\simeq 1.4\times 10^{-9}$. The figure shows that, provided the maximum ratio is larger than one, one generally obtains {\it two solutions}, one at low and one at high mass.

Our final Fig.~\ref{fig:yieldcontour} shows the two solutions (the lower-mass one on the left and the higher mass one on the right), on the parameter space defined by plane $(w,T_r)$, in a Log10 scale of the DM mass in GeV. The white regions in the upper and lower parts of the plot do not lead to a sufficient DM abundance. As shown in the figure, the viable masses producing the observed cosmological DM abundance span the extensive range from approximately 10 keV up to close to the assumed reheating scale.

\section{Discussion and Conclusions}\label{sec:conclusions}

We studied particle production from quasi-de Sitter cosmic horizons in the early universe. Our results rest on several key assumptions. First, we posit the existence of a post-inflationary accelerated expansion phase with an equation of state $P = w\rho$ with $-1 < w < -1/3$, ending at temperatures well above 10 MeV and transitioning to radiation domination before Big Bang Nucleosynthesis. Secondly, we adopt the static observer perspective, assuming that we can be considered static observers with respect to this early quasi-de Sitter phase. This assumption allows us to treat the cosmological horizon as a physical, stationary boundary emitting thermal radiation. We further assume that the cosmic horizon in this accelerating universe has a temperature inversely proportional to the proper cosmic horizon size. Our calculations are based on the frozen-in density of a stable particle of mass $m$ produced by this cosmic horizon, further assuming no number-changing processes occur in the late universe. 

The results presented in Section \ref{sec:production} demonstrate that particle production from quasi-de Sitter cosmic horizons in the early universe can provide a viable mechanism for generating the observed cosmological DM abundance, over a staggering range of masses. This scenario offers a novel approach to DM production that does not rely on traditional thermal freeze-out or freeze-in mechanisms, but instead leverages the fundamental properties of expanding spacetimes with cosmic horizons.

Our analysis reveals several key findings:

\begin{enumerate}
    \item The DM yield is sensitive to three main parameters: the DM particle mass $m$, the equation of state parameter $w$, and the transition temperature $T_r$ at which the universe transitions from the quasi-de Sitter phase to radiation domination;

\item For suitable choices of these parameters, there exist solutions that produce the correct DM abundance.  This is illustrated in Figure \ref{fig:yield}, which shows that for given values of $w$ and $T_r$, there are two mass solutions - one at lower mass and one at higher mass - that yield the observed cosmological DM density;

\item The viable parameter space, as shown in Fig.~\ref{fig:yield}, spans a significant range of $w$ and $T_r$ values;

\item The mass of DM particles produced through this mechanism can range from a few keV up to near the Planck scale, depending on the specific values of $w$ and $T_r$. A distinctive feature of this production mechanism is thus that it naturally accommodates both light as well as very heavy DM candidates.

\item The dependence on the reheating temperature $T_{reh}$ is relatively weak, as shown in Fig.
\ref{fig:limitintegrand}, particularly for lower values of $T_r$ and $w$ close to the de Sitter limit $w\to -1$. This indicates that the mechanism is relatively insensitive to the details of the inflationary period, adding to its robustness.

\end{enumerate}

From a particle physics perspective, the mechanism discussed here provides motivation for exploring very heavy DM candidates, well beyond the range typically considered in WIMP scenarios. Cosmologically, the mechanism additionally offers a new way to connect the properties of DM to the dynamics of the early universe. The dependence on the equation of state parameter $w$ provides a direct link between DM production and the nature of the energy content driving the quasi-de Sitter expansion phase. 

In sum, we showed that particle production from quasi-de Sitter cosmic horizons represents a  novel mechanism for DM production in the early universe. It naturally produces   DM candidates over a very broad range of masses and it establishes an intriguing connection between DM properties and early universe dynamics. While further work is needed to fully explore its implications and refine the model, this mechanism opens up new possibilities for understanding the nature and origin of DM.

\section*{Acknowledgements} \label{sec:acknowledgements}
   This material is based upon work supported in part by the U.S. Department of Energy grant number de-sc0010107. The Author is grateful to  Edgar Shaghoulian and  Batoul  Banihashemi for important and useful remarks, thoughts, and feedback.

\bibliographystyle{apsrev4-1}
\bibliography{references}
\end{document}